\documentclass[3p]{elsarticle}

\usepackage{graphicx,enumitem,multirow}
\usepackage{amsfonts,amsmath,amssymb,amsthm,bm}
\usepackage[bookmarks=false]{hyperref}

\journal{Journal of \LaTeX\ Templates}
\usepackage{color}

\biboptions{sort&compress}
\newcommand{\iid}{i.\@i.\@d.\@ }

\newcommand{\eg}{\emph{e.g.}, }
\newcommand{\ie}{\emph{i.e.}, }
\newcommand{\ud}{\mathrm{d}}

\renewcommand{\hat}{\widehat}
\renewcommand{\leq}{\leqslant}
\renewcommand{\geq}{\geqslant}

\newproof{pf}{Proof}
\newdefinition{rmk}{Remark}
\newtheorem{proposition}{Proposition}

\usepackage{geometry}

\begin{document}

\begin{frontmatter}

\title{Bounded Regression with Gaussian Process Projection}


\author{Jize Zhang}
\address{Department of Civil \& Environmental Engineering \& Earth Sciences, The University of Notre Dame}


\author{Lizhen Lin }
\address{Department of Applied and Computational Mathematics and Statistics, The University of Notre Dame}

\begin{abstract}
Examples with bound information on the regression function and density abound in many real applications. We propose a novel approach for estimating such functions by incorporating the prior knowledge on the bounds. Specially, a Gaussian process is first imposed on the regression function whose posterior distribution is then projected onto the bounded space.  The resulting projected measure is then used for inference. The projected sample path has closed form which facilitates  efficient computations. In particular,  our projection approach maintains a comparable computational efficiency with that of the original GP. The proposed method yield predictions that respects  bound constraints everywhere, while allows varying bounds across the input domain.  An extensive simulation study is carried out which demonstrates that the performance of our approach dominates that of the competitors. An application to real data set is also considered. 
\end{abstract}


\end{frontmatter}

\section{Introduction}\label{sec:intro}
In many interesting applications of the regression problem, prior knowledge is often available on the lower and/or upper bound of responses. Examples range from non-negative constraints on physical quantities,  interval restrictions on proportion measures, to constraining probability density with bounds. Incorporating  such bounds in the estimation process can often substantially improve the inference in terms of efficiency and stability, while also enforcing the prediction to honor such prior knowledge.


In recent decades, Gaussian processes (GPs) have emerged  as a very powerful tool in statistics and machine learning  due to their flexibility in interpolation and regression, and  unique ability for modeling prediction uncertainty \citep{rasmussen_gaussian_2005}.  However, GPs' prediction and uncertainty quantification by definition are unbounded. 
 To incorporate bound information with GPs  in a most intuitive manner,  one can perhaps rely on  truncated GPs analogous to the truncated Gaussian distribution \cite{jensen2013bounded}. However, this  truncated version loses the analytical tractability of the original GP,  consequently both inference and prediction have to be carried out via approximation methods such as Expectation Propagation. The Soft Kriging approach \cite{journel1986constrained} modifies the covariance models to satisfy bound constraints, but the complicated re-formulation of the covariance models can be difficult to derive in practice \cite{yoo2006area}. Another related approach is beta regression \cite{ferrari2004beta}, which makes use of the beta distribution with a bounded support. However, beta regression  is mostly suited for applications with fixed ranged bounds.

Alternatively, the bounded response can be transformed into a unbounded domain that can be better modeled by a GP. The most common approach of this type is to use logarithm transformation to predict non-negative responses, \eg the probability density function \cite{rasmussen2003gaussian}. More sophisticated transformations are also available and discussed in \cite{jensen2013bounded,michalak2008gibbs}. They share a similar drawback to that of the beta regression: most of them are unsuitable for variable bounds.

Stochastic sampling based approaches have also been proposed to address various types of shape constraints in GPs, including inequality or equality \cite{freulon1993conditioning,abrahamsen2001kriging, michalak2008gibbs,maatouk2017finite}, monotonicity \cite{kleijnen2013monotonicity, riihimaki2010gaussian,wang2016estimating,golchi2015monotone,maatouk2017finite}, and convexity \cite{wang2016estimating,maatouk2017finite}. The general idea is to simulate posterior GP sample paths satisfying constraints at pre-specified input locations. This framework  has the advantage of directly accommodating  various type of shape constraints, but suffers from the shortcomings that the GP's computational efficiency and numerical stability are in question  when  additional conditioning on input locations increases affecting the numerical complexity of the problem \cite{riihimaki2010gaussian, wang2016estimating, golchi2015monotone}. Furthermore, the constraints are typically only met at the chosen input locations instead of the whole domain. Increasing the number of conditioned input locations can compensate for the loss in efficency, but comes at the cost of further exacerbating the aforementioned computational challenge. Consequently, the application of such approaches to higher-dimensional problems has been relatively limited.

Similarly but in a deterministic manner, Lagrange-multiplier based approaches have been exploited \cite{barnes1992adding} Michaelka 2003. The use of Lagrange multipliers leads to the `best' prediction honoring the constraints according to the original GP's likelihood. However, such methods lack a probabilistic interpretation, hence cannot be used to generate uncertainty prediction, confidence interval, or conditional realizations. The computation can be also challenging if the number of inputs to be predicted is high.

Both conditional sampling and Lagrange multiplier methods are \emph{a posteriori} approaches, in the sense that they all postprocess the original GP by adding some \emph{a posteriori} constraints to enforce the GP's prediction or conditional realizations to incorperate the bound information. Our work proposes another a posteriori strategy by projecting the GP posterior onto the bounded space. The idea follows the line of work proposed in \cite{lin2014bayesian}, which projects the GP posterior sample paths onto the monotone constrained space. 
The monotonic projection was approximated via the pooled adjacent violators (PAV) algorithm. 
Under this projection framework, we present the methodology to incorporate the bound constraints into GP. Unlike in \cite{lin2014bayesian}, the projection of sample paths can be analytically carried out for bounded constraints. As a result, our projection approach maintains a comparable computational efficiency with that of the original GP, and produces closed-form expression for the bounded posterior prediction moments and distribution functions. The proposed method yield predictions that honor bound constraints everywhere, allows varying bounds across the input domain. We also propose a GP parameter inference algorithm that can take the bound information into account at the inference stage.

The paper is organized as follows. 
Section \ref{sec:bgp} presents the projection approach to enforce the bound constraints, as well as the prediction formulas by projecting the posterior GP.  Section \ref{sec:inf} discusses an computational implementation to incorporate the bound information in the GP parameter inference procedure.  The improvement brought by the proposed strategies are discussed in Section \ref{sec:exa} with both synthetic-data and real-data problems. Concluding remarks are given in Section \ref{sec:con}.

\section{Bounded Gaussian Process (bGP)}\label{sec:bgp}
\subsection{A Brief Overview of Gaussian Processes}\label{sec:gp}

A Gaussian Process (GP) is distribution for random functions whose any finite subset evaluations have a joint Gaussian distribution. Consider the $n_x$-dimensional input vector $\mathbf{x} \in \mathcal{X} \subset \mathbb{R}^{n_x}$ and the output function $y=f(\mathbf{x}):\mathcal{X} \rightarrow \mathbb{R}$. We model $f(x)$ by the GP \cite{sacks1989design}: $f(\mathbf{x}) \sim \mathcal{GP}\left(m(\mathbf{x}),k(\mathbf{x},\mathbf{x}')\right)$, which is equivalent to placing a GP prior distribution over the class of output functions. For the ease of illustration, we adopt a zero-mean Gaussian process prior (\ie $m(\mathbf{x})=0$), with the squared exponential covariance function:
\begin{align}
k(\mathbf{x},\mathbf{x}') = 	\sigma^2\exp \left(\sum_{i=1}^{n_x}-\frac{(x_i-x_i')^2}{2\theta_i^2}\right)
\label{corr}
\end{align}
where $\sigma^2$ and $\theta = (\theta_1,\ldots,\theta_{n_x})$ are GP parameters to be inferred. In general any popular covariance functions (\eg Generalized exponential, Ornstein-Uhlenbeck, or Mat\'{e}rn ones) can be used, as long as the GP sample paths are continuous in a $\mathcal{C}_0$ sense.

Suppose that we obtain $N$ observations of the output $\mathbf{Y}=[\mathbf{y}_1,\ldots,\mathbf{y}_N]$ at the set of inputs $\mathbf{X} = [\mathbf{x}_1,\ldots,\mathbf{x}_N]$. The collection of observations $\mathbf{Y}$ will then have the following Gaussian distributions:
\begin{align}
\label{eq-model}
\mathbf{Y} \sim \mathcal{N}\left(0,\mathbf{K}\right)
\end{align}
where $\mathbf{K}$ is a $N\times N$ square matrix whose element $[\mathbf{K}]_{ij}$ is given by $k(\mathbf{x}_i,\mathbf{x}_j)$ in Eq.\eqref{corr}.

Under the specified GP prior, conditioned on the data $\mathbf{X}$ and $\mathbf{Y}$ as well as the GP parameters $(\sigma^2,\theta)$, the prediction (posterior) distribution of $f(.)$ at an arbitrary input location $\mathbf{x}^*$ will be a Gaussian distribution: 
\begin{align}
f(\mathbf{x}^*)\sim \mathcal{N}\left(\mu_f(\mathbf{x}^*),\sigma^2_f(\mathbf{x}^*)\right)
\label{gpdist}	
\end{align}
with the following mean and variance expression:
\begin{align}
\begin{split}
\mu_f(\mathbf{x}^*) =& \mathbf{k(x^*)}^T\mathbf{K}^{-1}\mathbf{Y}\\
\sigma^2_f(\mathbf{x}^*)=& \sigma^2-\mathbf{k(x^*)}^T\mathbf{K}^{-1}\mathbf{k(x^*)}
\end{split}
\label{gpmeanvar}
\end{align}
where $\mathbf{k(x^*)}$ is the $N\times 1$ dimensional vector whose entries are output covariances between $\mathbf{x}^*$ and $N$ design points $\mathbf{X}$, obtained by $[\mathbf{k}(\mathbf{x}^*)]_j = k(\mathbf{x}^*,\mathbf{x}_j)$.

If some information is available in the form of lower or upper bounds for the output, it would be desirable to incorporate such bound information into the GP prediction. The potential benefit comes in multifolds: (a) incorporating the bound information often improves the efficiency of estimates; (b) the bound information might be able to complement the sparsity of training data at certain input locations, therefore accounting for them should enhance the prediction accuracy; and (c) the Gaussian nature makes the posterior prediction distribution in Eq.\eqref{gpdist} unbounded, and  incorporating the constraints by projection produces realistic results. Therefore, accounting for bound information will be crucial for inference.

In this paper, we assume that constraints are known beforehand and can be expressed by a continuous lower-bound function $l(\mathbf{x}): \mathcal{X} \rightarrow \mathbb{R}$, and/or a continuous upper-bound function $u(\mathbf{x}): \mathcal{X} \rightarrow \mathbb{R}$. We then propose the \emph{Bounded Gaussian Process (bGP)}, an approach to enforce bound constraints by projecting the unconstraint GP sample paths onto the bounded space, which results in closed-formed bounded posterior distributions.

\begin{rmk}
The model in \eqref{eq-model} can easily incorporate some noise in the observing $Y$ by adding the variance of the noise to the diagonal of matrix $K$. 
\end{rmk}

\subsection{Projection to the bounded space}
Let $\tilde{f}$ be a GP sample path. Let $\mathcal{B}$ represent the subset of the space of continuous functions satisfying the bound constraints:
\begin{align*}
\mathcal{B}=\left\{ f \in \mathcal{C}^0(\mathcal{X}): -\infty < l(\mathbf{x})\leq f(\mathbf{x}) \leq u(\mathbf{x}) < \infty, \mathbf{x} \in \mathcal{X} \right\}
\end{align*}
We define the projection of the unbounded GP sample path to the bounded continuous function space $\mathcal{B}$ as follows:
\begin{align}
\tilde{g} = \text{argmin}_{g \in \mathcal{B}} \int_{\mathcal{X}} \left\{g(x)-\tilde{f}(x)\right\}^2 \ud x
  \label{proj}
\end{align}
The projection minimizes the integrated squared error.

\begin{figure}
\centering
\includegraphics[width=20pc]{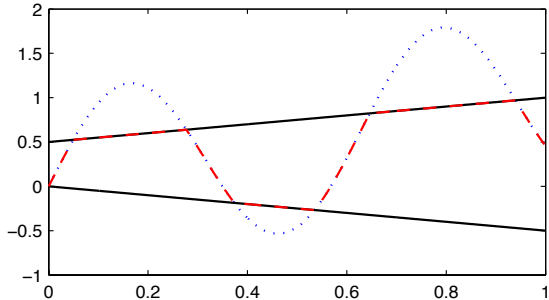}	
\caption{Example of a GP sample path $\tilde{f}$ (blue dashed lines) and its projection $\tilde{g}$ (red dash lines). Bounds are shown in black solid lines}
\label{proj_fig}
\end{figure}

For each $\tilde{f}$, its projection $\tilde{g}$ has a unique solution as follows:
\begin{align}
\tilde{g}(\mathbf{x}) = \begin{cases}
l(\mathbf{x})& \tilde{f}(\mathbf{x})\leq l(\mathbf{x})\\
\tilde{f}(\mathbf{x})& l(\mathbf{x})<\tilde{f}(\mathbf{x})<u(\mathbf{x})\\
u(\mathbf{x})& \tilde{f}(\mathbf{x})\geq u(\mathbf{x})
\end{cases}
\label{projsol}
\end{align}
Figure \ref{proj_fig} illustrates the idea of such projection. 

The proof for the existence and uniqueness of the projection solution follows.
\begin{proposition}
The projection $\tilde g$ defined in \eqref{projsol} exists and is unique. 
\end{proposition}
\begin{pf}
The existence of $\tilde{g}$ is trivial to prove if $\tilde{f} \in \mathcal{B}$, when $\tilde{f} \equiv \tilde{g}$. If $\tilde{f} \notin \mathcal{B}$, note that $l,u$ and $\tilde{f}$ are all continuous in $\mathcal{X}$. According to Theorem 4.2.1 in  \cite{strichartz2000way}, the projection $\tilde{g}$ defined in Eq.\eqref{projsol} is also continuous in $\mathcal{X}$. Notice that $\forall \mathbf{x} \in \mathcal{X}$, $l(\mathbf{x})\leq \tilde{g}(\mathbf{x}) \leq u(\mathbf{x})$, $\tilde{g}$ then indeed belongs to the bounded continuous subset $\mathcal{B}$. 

For  uniqueness, we assume by contrary that there were another solution $\tilde{g}' \neq \tilde{g}$ minimizing the integrated squared error described in Eq.\eqref{proj}. We further partition the entire domain $\mathcal{X}$ into 3 sub-divisions, $\mathcal{X}_L= \{\mathbf{x} \in \mathcal{X}|l(\mathbf{x}) \leq \tilde{f}(\mathbf{x})\}$, $\mathcal{X}_U= \{\mathbf{x} \in \mathcal{X}|u(\mathbf{x}) \geq \tilde{f}(\mathbf{x})\}$, and also $\mathcal{X}_M= \{\mathbf{x} \in \mathcal{X}|l(\mathbf{x}) < \tilde{f}(\mathbf{x})< u(\mathbf{x})\}$. We can observe that
\begin{align}
\begin{split}
 &\int_{\mathcal{X}} \left\{\tilde{g}'(\mathbf{x})-\tilde{f}(\mathbf{x})\right\}^2 \ud \mathbf{x}-\int_{\mathcal{X}} \left\{\tilde{g}(\mathbf{x})-\tilde{f}(\mathbf{x})\right\}^2 \ud \mathbf{x}\\
 =&\int_{\mathcal{X}_M} \left\{\tilde{g}'(\mathbf{x})-\tilde{f}(\mathbf{x})\right\}^2 \ud \mathbf{x}
 +\int_{\mathcal{X}_L}\left\{\left(\tilde{g}'(\mathbf{x})-l(\mathbf{x})\right)\left(\tilde{g}'(\mathbf{x})+l(\mathbf{x})-2\tilde{f}(\mathbf{x})\right)\right\}\ud \mathbf{x}\\
 +&\int_{\mathcal{X}_U} \left\{\left(\tilde{g}'(\mathbf{x})-u(\mathbf{x})\right)\left(\tilde{g}'(\mathbf{x})+u(\mathbf{x})-2\tilde{f}(\mathbf{x})\right)\right\} \ud \mathbf{x}.\\
 \end{split}
\label{dist}
\end{align}

As $\tilde{g}'\in \mathcal{B}$, according to the definition of $\mathcal{B}$ and the sub-divisions we have
\begin{align}
\begin{split}
&\tilde{g}'(\mathbf{x})-l(\mathbf{x})\geq 0, \tilde{g}'(\mathbf{x})+l(\mathbf{x})-2\tilde{f}(\mathbf{x})\geq 0, \forall \mathbf{x} \in \mathcal{X}_L \\
\Rightarrow &\int_{\mathcal{X}_L}\left\{\left(\tilde{g}'(\mathbf{x})-l(\mathbf{x})\right)\left(\tilde{g}'(\mathbf{x})+l(\mathbf{x})-2\tilde{f}(\mathbf{x})\right)\right\}\ud \mathbf{x} \geq 0\\
&\tilde{g}'(\mathbf{x})-u(\mathbf{x})\leq 0, \tilde{g}'(\mathbf{x})+u(\mathbf{x})-2\tilde{f}(\mathbf{x})\leq 0, \forall \mathbf{x} \in \mathcal{X}_U\\
\Rightarrow &\int_{\mathcal{X}_U} \left\{\left(\tilde{g}'(\mathbf{x})-u(\mathbf{x})\right)\left(\tilde{g}'(\mathbf{x})+u(\mathbf{x})-2\tilde{f}(\mathbf{x})\right)\right\} \ud \mathbf{x}\geq 0	.
\end{split}
\label{dist2}
\end{align}

Applying Eq.\eqref{dist2} to Eq.\eqref{dist}, we find that the integrated squared error of $\tilde{g}'$ is always greater or equal to that of $\tilde{g}$, where the equality takes if and only if $\tilde{g}'=\tilde{g}$ everywhere. Hence, the uniqueness of $\tilde{g}$ follows. 

\end{pf}

\begin{rmk}
The projection is surjective, since $\tilde{g}(\mathbf{x})=\tilde{f}(\mathbf{x})$ if $\tilde{f} \in \mathcal{B}$. Therefore, such projection of GP sample paths introduces a valid measure on the set of bounded and continuous functions $\mathcal{B}$.	
\end{rmk}

\subsection{Inference by projecting the posterior distribution}

Bounded posterior inference can be carried out by projecting sample paths from an unconstraint GP prior or from the posterior. The former one is in a fully Bayesian sense, but the additions of bounded constraints makes it more challenging to design efficient MCMC samplers for drawing realizations from the now non-GP posterior. Alternatively, the projection of the GP posterior has been shown to correspond to an empirical Bayes approach \cite{lin2014bayesian}, but has the great  advantage it leads to easy and straightforward computations of posterior distribution functions and moments. Therefore, we only focus on the latter approach in this paper.  

For an arbitrary input $\mathbf{x}^*$ with the posterior output distribution as $f(\mathbf{x}^*)\sim \mathcal{N}(\mu_f(\mathbf{x}^*),\sigma^2_f(\mathbf{x}^*))$, the cumulative distribution function (CDF) for its projected output $\tilde{g}(\mathbf{x}^*)$ in Eq.\eqref{projsol} can be written as:

\begin{align}
\mathbb{F}\left(\tilde{g}(\mathbf{x}^*)\right) = 
\begin{cases}
\Phi\left(l(\mathbf{x}^*);\mu_f(\mathbf{x}^*),\sigma^2_f(\mathbf{x}^*)\right) & \tilde{g}(\mathbf{x}^*)= l(\mathbf{x}^*)\\
\Phi\left(\tilde{g}(\mathbf{x}^*;\mu_f(\mathbf{x}^*),\sigma^2_f(\mathbf{x}^*)\right) & l(\mathbf{x}^*)<\tilde{g}(\mathbf{x}^*)<u(\mathbf{x}^*)\\
1 & \tilde{g}(\mathbf{x}^*)= u(\mathbf{x}^*)
\end{cases}
\label{probc}
\end{align}

while its probability density function\footnote{The probability measure $\mathbb{P}$ is defined over the measurable space ([$l(\mathbf{x}), u(\mathbf{x})$], $\mathfrak{B}$) where $\mathfrak{B}$ is the class of all Borelian subsets of [$l(\mathbf{x}), u(\mathbf{x})$], is such that $\mathbb{P} << \lambda + \delta_{l(x)} + \delta_{u(x)}$, with $\lambda$ representing the Lebesgue measure and $\delta_c$ is a point mass at $c$, \ie $\delta_c(A) = 1$, if $c \in A$ and 0, if $c \notin A$, $A \in \mathfrak{B}$.} (PDF) is:

\begin{align}
\mathbb{P}\left(\tilde{g}(\mathbf{x}^*)\right) = \begin{cases}
\Phi\left(l(\mathbf{x}^*);\mu_f(\mathbf{x}^*),\sigma^2_f(\mathbf{x}^*)\right) & \tilde{g}(\mathbf{x}^*)= l(\mathbf{x}^*)\\
\phi\left(\tilde{g}(\mathbf{x}^*);\mu_f(\mathbf{x}^*),\sigma^2_f(\mathbf{x}^*)\right) & l(\mathbf{x}^*)<\tilde{g}(\mathbf{x}^*)<u(\mathbf{x}^*)\\
1-\Phi\left(u(\mathbf{x}^*);\mu_f(\mathbf{x}^*),\sigma^2_f(\mathbf{x}^*)\right) & \tilde{g}(\mathbf{x}^*)= u(\mathbf{x}^*)
\end{cases}
\label{probg}
\end{align}
where $\Phi(.;a,b)$ and $\phi(,;a,b)$ represent the cumulative/probability density function for the normal distribution with mean equal to $a$ and variance $b$ respectively. 
Note that the probability of observing the projection at the lower or upper bounds appearing in Eq.\eqref{probg} are the probability masses. Figure \ref{dis_fig} presents such distribution for the values of $(\mu_f = 0, \sigma_f = 1, l = -1, u=2)$. 
\begin{figure}
\centering
\includegraphics[width=20pc]{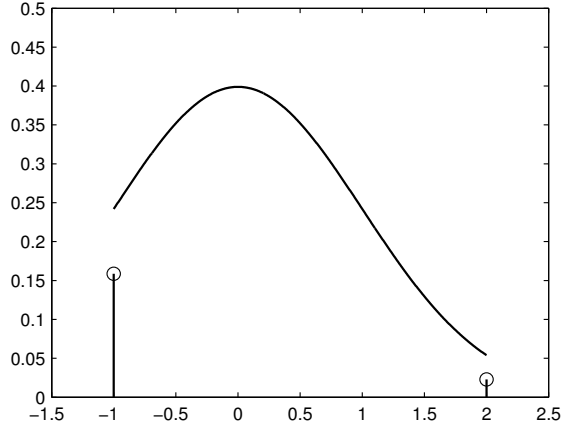}	
\caption{Example of the distribution function for $(\mu_f = 0, \sigma_f = 1, l = -1, u=2)$. The solid line represents probability density functions and the circle-solid-lines are the probability masses at the bounds}
\label{dis_fig}
\end{figure}

This distribution in Eq.\eqref{probc}-\eqref{probg} can be seen as a mixture of two distributions: a discrete two-point distribution on the bounds, and a continuous truncated Gaussian distribution within the bounds. Let's temporarily drop the dependency on $\mathbf{x}^*$ to avoid notational clutter. The mixture distribution can be expressed as:
\begin{align*}
\mathbb{F}\left(\tilde{g}\right) = \left(1-Z\right)\cdot\text{Ber}\left(\frac{\tilde{g}-l}{u-l};\gamma\right)+Z\cdot \Psi\left(\tilde{g};\mu_f,\sigma^2_f,l,u\right)
\end{align*}
where the mixture parameter for the two distributions $Z$ is given by:
\begin{align*}
Z=\Phi\left(u;\mu_f,\sigma^2_f\right)-\Phi\left(l;\mu_f,\sigma^2_f\right)	
\end{align*}
Here Ber denotes the CDF function of a Bernoulli random variable with the parameter $\gamma = [1-\Phi\left(\beta\right)]/(1-Z)$, and $\Psi$ represents the CDF function of a truncated Gaussian distribution defined on the interval $[l,u]$, whose non-truncated version has the mean $\mu_f$ and variance $\sigma_f^2$.

To calculate the mean and variance of the projected posterior, we define the following terms to maintain the notational clarity:
\begin{align*}
\alpha \triangleq \left(l-\mu_f\right)/\sigma_f,\beta \triangleq  \left(u-\mu_f\right)/\sigma_f	
\end{align*}
After some algebra, the projected posterior mean can be computed in a closed-form formula:
\begin{align}
\mu_g = Z\cdot \mu_f+\left[\phi\left(\alpha;0,1\right)-\phi\left(\beta;0,1\right)\right]\cdot \sigma_f+l\cdot\Phi\left(\alpha;0,1\right)+u\cdot[1-\Phi\left(\beta;0,1\right)]
\label{predmean}
\end{align}
%
The variance of the projection can be computed as:
\begin{align}
\begin{split}
\sigma_g^2 =& Z\cdot\left[\sigma_f^2+\mu_f^2\right]+2\mu_f\sigma_f\cdot \left[\phi\left(\alpha;0,1\right)-\phi\left(\beta;0,1\right)\right]+\sigma^2_f \cdot \left[ \alpha \phi\left(\alpha;0,1\right)-\beta\phi\left(\beta;0,1\right) \right]\\+& l^2\cdot \Phi\left(\alpha;0,1\right)+u^2\cdot [1-\Phi\left(\beta;0,1\right)]-\mu_g^2
\end{split}
\label{predvar}
\end{align}

\begin{rmk}
\label{rmk2}
One is able to relax the continuous bound assumptions and apply the formulas to problems with discontinuous bounds. The empirical Bayes interpretation will be missing, but one can simply rely on the formulas in this section as a correction formula to enforce the bound requirements. 
\end{rmk}

\begin{rmk}
\label{rmk3}
We provide the posterior projection mean and variance for the one-sided case that only the lower bound $l$ exists (the upper-bound only cases are analogous) as follows:
\begin{align}
\begin{split}
\mu_g =& Z\cdot \mu_f+\phi\left(\alpha;0,1\right)\cdot \sigma_f+l\cdot\Phi\left(\alpha;0,1\right)\\
\sigma_g^2 =& Z\cdot\left[\sigma_f^2+\mu_f^2\right]+\left[2\mu_f\sigma_f+\sigma^2_f \cdot  \alpha \right] \cdot \phi\left(\alpha;0,1\right)+ l^2\cdot \Phi\left(\alpha;0,1\right)-\mu_g^2
\end{split}
\label{predmeanvar1}
\end{align}
%
with $Z = 1-\Phi\left(l;\mu_f,\sigma^2_f\right)$ in this case. 
\end{rmk}

\section{GP parameter inference with the incorporation of bound information}\label{sec:inf}

It remains to discuss about the selection of parameters ($\sigma^2,\theta$) appearing in Eq.\eqref{corr}. Instead of conducting a fully Bayesian analysis and integrating them out, we rely on a point estimate sought by optimization for simplicity. Two popular approaches are often used for this task \cite{bachoc2013cross}, namely the \emph{maximum-likelihood} estimation or a \emph{cross-validation} approach. Fully ignoring the bound information, GP parameters can be readily inferred basing on either approach. Nonetheless, our goal is to select the GP parameters basing also on the bound information, which naturally requires working with the projections $\tilde{g}$ instead of the original GPs $\tilde{f}$. For $\tilde{g}$, the likelihood principle would be inapplicable, due to its mixed discrete-continuous distribution nature. We thus turn to cross-validation based approaches to estimate the parameters. 

\subsection{Cross-validation GP parameter inference without bounds}
\label{sec:inf1}
Let us first recall that for the unconstraint GP, the cross-validation procedure to determine ($\sigma^2,\theta$) involves the two following steps \cite{bachoc2013cross}:

\begin{enumerate}
\item Determine $\theta$ by solving the following minimization problem:
\begin{align}
\hat{\theta} = \text{argmin}_{\theta} \left\{\sum_{i=1}^{N} \left(y_i - \mu_{f,-i} \right)^2\right\}
\end{align}
where $\mu_{f,-i} $ is the leave-one-out-cross-validation (LOO-CV) predictor, \ie the posterior mean of data sample $x_i$ based on all training data except itself. It can be computed analytically:
\begin{align}
\mu_{f,-i} =& y_i-\left[\mathbf{K}^{-1}\mathbf{Y}\right]_i/\left[\mathbf{K}^{-1}\right]_{ii}
\label{predloo1}
\end{align}
Thus the resulting estimator $\theta$ essentially minimizes LOO-CV predicted residual sum-of-squares (PRESS).
\item Notice that the PRESS criterion does not depend on the variance parameter $\sigma^2$, an additional step has to be carried out to calibrate it. According to \cite{cressie2015statistics}, the average ratio between the squared LOOCV errors $(y_i-\mu_{f,-i})^2$ and the LOO-CV variance of data samples $\sigma_{f,-i}^2$ should be optimized to be close to one, where \begin{align}
\sigma^2_{f,-i}=& 1/\left[\mathbf{K}^{-1}\right]_{ii}
\label{predloo2}
\end{align}
is the LOO-CV variance for the $i$-th training sample. The following closed-form expression exists for such variance parameter estimation:
\begin{align}
\hat{\sigma}^2 = \frac{1}{N}\mathbf{Y}^T\mathbf{K}^{-1}[\text{diag}(\mathbf{K}^{-1})]^{-1}\mathbf{K}^{-1}\mathbf{Y}
\end{align}	
with diag(.) being the matrix obtained by setting to 0 all non-diagonal terms of the one inside the parenthis.
\end{enumerate}

\subsection{Cross-validation GP parameter inference with bounds}
\label{sec:inf2}
We propose to calibrate the GP parameters in a similar manner to the unconstraint case, by minimizing the LOO-CV PRESS. The difference is that we replace the predictor from the unconstraint LOO-CV mean $\mu_{f,-i}$ to the LOO-CV projected mean $\mu_{g,-i}$, which can be obtained by plugging the LOO-CV mean and variance $\mu_{f,-i}$ and $\sigma^2_{f,-i}$ back into Eq.\eqref{predmean}-\eqref{predvar}: 
\begin{align}
(\hat{\sigma}^2,\hat{\theta}) = \text{argmin}_{\sigma^2,\theta} \left\{\sum_{i=1}^{N} \left( y_i - \mu_{g,-i} \right)^2\right\}
\label{press}
\end{align}
As $\mu_{g,-i}$ also depends the variance parameter $\sigma^2$, the PRESS criterion after projection has one more degree of freedom in contrast to its unconstraint counterpart, and would not necessarily require a separate calibration step for $\sigma^2$. However, it is important to point out that PRESS's sensitivity with respect to $\sigma^2$ could be low for certain extreme circumstances, \eg when all lower and upper bounds for the training data approach $-\infty$ and $\infty$ respectively, the dependency of PRESS in Eq.\eqref{press} on $\sigma^2$ will be negligble. To mitigate such problem, we impose an additional constraint for the variance estimator, restricting it to stay in a reasonable range, which is established based on the unconstraint LOOCV variance estimator in Eq.\eqref{predloo2}, specifically:  

\begin{equation}
\begin{aligned}
& \underset{\sigma^2,\theta}{\text{argmin}}
& & \left\{\sum_{i=1}^{N} \left( y_i - \mu_{g,-i} \right)^2\right\} \\
& \text{subject to}
& & \sigma^2  \geq c_l(\frac{1}{N}\mathbf{Y}^T\mathbf{K}^{-1}[\text{diag}(\mathbf{K}^{-1})]^{-1}\mathbf{K}^{-1}\mathbf{Y})  \\
& & & \sigma^2 \leq c_u(\frac{1}{N}\mathbf{Y}^T\mathbf{K}^{-1}[\text{diag}(\mathbf{K}^{-1})]^{-1}\mathbf{K}^{-1}\mathbf{Y}).
\end{aligned}
\label{projcon}
\end{equation}

For $c_l$ and $c_u$ a suggested pair of values are ($10^{-2},10^{2}$), which ensures the estimated variance parameter is no more than two order of magnitude different from the unconstraint one.

Maximizing over the parameters is highly non-trivial since the objective function can be non-convex and multimodal. We thus resort to the covariance matrix adaptation evolution strategies (CMA-ES) \cite{hansen2001completely} to seek for the optimizer in Eq.\eqref{projcon}. The potential gain of incorporating the bound information at the inference stage will be further discussed in Section 5. 

\section{Illustrative examples}\label{sec:exa}

 In this section, we demonstrate the behavior of the proposed methods and the potential gain from incorporating bound constraints in prediction and inference stages, through several examples with different dimensions. The methodology is applied in three simulated examples and one real data application.  Comparisons are carried out between our method and some prominent competitors. 

\subsection{Synthetic examples}
\subsubsection{1-D examples}
Consider first synthetic one-dimensional problems with the following output functions and bounds proposed by \cite{papp2014shape} and \cite{da2012gaussian}, as shown in Figure \ref{dis_fig}:

\begin{enumerate}[label=(\alph*)]
\item $f(x) = \frac{1}{5} \mathcal{B}_{1.4,2.6} \left( \frac{x-3}{5}\right), x \in [0,10] $; with the constant lower bound $l(x) = 0$. Here $\mathcal{B}_{\alpha,\beta}$ denotes the beta PDF with parameters $\alpha$ and $\beta$.
\item $f(x) = x^2\sin(x^{-1}), x \in [-\frac{\pi}{8},\frac{\pi}{8}]$; with bounds $l(x)=-x^2,u(x)=x^2$;
\item $f(x) = \frac{\sin(10\pi x^{5/2})}{10\pi x}, x \in [0,1]$; with a single lower or upper bound to denote the sign of $f(x)$ as
\begin{align*}
\begin{cases}
  l(x)=0, & \text{if } f(x) \geq 0, \\
  u(x)=0, & \text{otherwise}.
\end{cases}
\end{align*}
\end{enumerate}
 
\begin{figure}
\centering
\includegraphics[width=20pc]{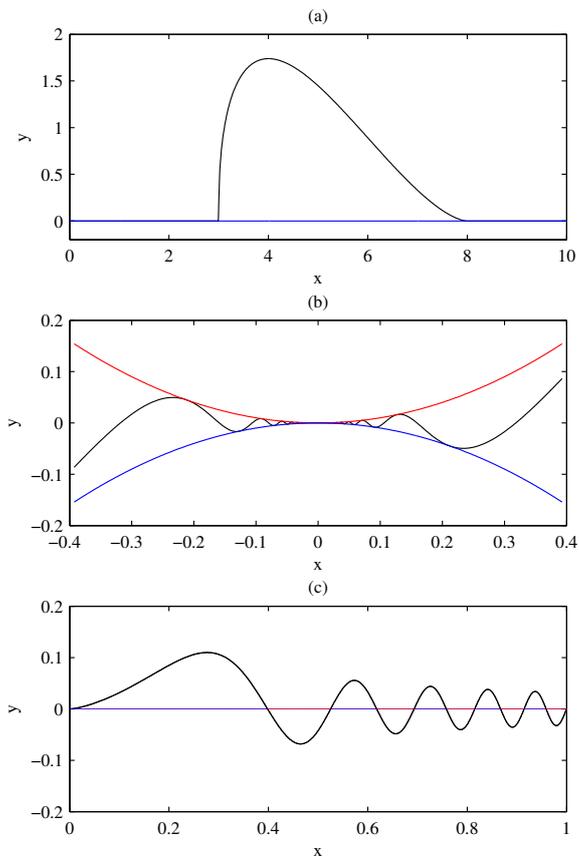}	
\caption{The output functions (black lines), and the lower (blue lines) and upper (red lines) bounds for all examined 1-D problems}
\label{dis_fig}
\end{figure}
 
These functions represent problems with different characteristics. The output in the first problem exhibits mild nonlinearities: it peaks in the middle and has a flat region at the beginning and the end. It is only assumed of an non-negative lower bound, mimicking the scenario of approximating a probability density function. The second one are more challenging to predict because of the strong nonlinearities introduced by the sine function and the frequent variations around $x=0$. It is equipped with varying lower and upper bounds to verify the applicability of the proposed method. The third one, proposed in \cite{da2012gaussian}, displays a non-stationary behavior and is thus also difficult to approximate. The bound functions for such problem only enforces the sign of the output, and thus are  discontinuous. As mentioned in Remark \ref{rmk3}, we can relax the modeling assumption on bounds, and use the prediction formulas based on projections simply as heuristic correction formulas. 

To investigate the overall performance and individual contribution of each component, we compare the proposed algorithm against several variants as well as an existing benchmark algorithm:

\begin{itemize}
\item \textbf{bGP}: the proposed approach, whose GP parameter is inferred \emph{incorporating} the bound information as in Section \ref{sec:inf2}, and the prediction is made \emph{with} projection as in Eq.\eqref{predmean}.
\item \textbf{bGP-I}: the bGP variant using only the parameter inference strategy. GP parameter is inferred \emph{incorporating} the bound information in Section \ref{sec:inf2}, but the prediction is made \emph{without} projection as in Eq.\eqref{gpmeanvar}.
\item \textbf{bGP-P}: the bGP variant using only the posterior projection strategy. In GP parameter inference, the bound information is neglected, but the prediction is made \emph{with} projection as in Eq.\eqref{predmean}. 
\item \textbf{GP}: the original GP approach, \ie bounds are \emph{neglected} in either the inference or prediction. 
\item \textbf{DM}: the comparison benchmark approach proposed by \cite{da2012gaussian}. GP parameter is inferred \emph{neglecting} the bound information as described in Section \ref{sec:inf1}, and the prediction is computed by the correlation-free conditional expectation formula given in Eq. (3.4) from \cite{da2012gaussian}. 
\end{itemize}

Fifty independent simulation trials are carried out. For each simulation, the training input locations are taken to be $N = 10$ (in problem (a) and (c)) or 15 (in problem (b)) Latin Hypercube sampling (LHS) points in the corresponding interval. For numerical convenience, we also pre-normalize the training input and output to have mean of zero and standard deviation of one. The prediction performance will be evaluated at $N_t = 1000$ test points equally spaced over the same interval. Three accuracy measures are employed, including the prediction coefficient of determination ($R^2$),
\begin{align*}
R^2 = 1-\frac{\sum_{i=1}^{N_t} \left(f(x^i)- \hat{f}(x^i) \right)^2}{\sum_{i=1}^{N_t} \left(f(x^i)- \frac{1}{N_t}\sum_{i=1}^{N_t}f(x^i) \right)^2}	
\end{align*}
 and the root mean squared error (RMSE):
\begin{align*}
RMSE = \sqrt{\frac{1}{N_t}\sum_{i=1}^{N_t} \left(f(x^i)- \hat{f}(x^i) \right)^2}
\end{align*}
as well as the average $95\%$ coverage probability over all test points:
\begin{align*}
CP = \frac{1}{N_t}\sum_{i=1}^{N_t}\mathbb{I}\{f(x^i)\in [Q_{0.975}^i,Q_{0.025}^i]\}
\end{align*}
where $\hat{f}$ is the predictor calculated from Eq.\eqref{gpmeanvar} (for bGP-I or GP) or Eq.\eqref{predmean} (for bGP or bGP-P), or by the correlation-free conditional expectation formula in \cite{da2012gaussian} (for DM). $Q_j^i$ is $j$-th posterior CDF for the $i$-th test sample. Note that the $CP$ measure is inapplicable to the DM approach, since no explicit formula for the posterior distribution function is given \cite{da2012gaussian}. Results on the accuracy measures are summarized in Table \ref{dens}-\ref{daviga}. 


\begin{table}
\centering
\begin{tabular}{ c c c c c c}
\hline
Problem (a)  & bGP & bGP-I & bGP-P & GP & DM \\ 
 \hline
 $R^2$ & \bf{96.1 $\pm$ 1.83} & 93.7 $\pm$ 3.98 & 95.6 $\pm$ 2.07 & 93.9 $\pm$ 3.73  & 94.6 $\pm$ 3.10 \\
 $RMSE$ & \bf{12.6 $\pm$ 2.99} & 16.0 $\pm$ 4.67 & 13.7 $\pm$ 3.12 & 15.7 $\pm$ 4.47 & 15.9 $\pm$ 4.11 \\
 $CP (\%)$ & 72.3 $\pm$ 14.0 & 72.3 $\pm$ 14.0 & \bf{85.7 $\pm$ 3.99} & \bf{85.7 $\pm$ 3.99} & N/A \\
\hline
\end{tabular}
\caption{$R^2$, $RMSE$ (both value multiplied by 100) and $CP(\%)$ of examined methods for Problem (a). The reported values are the average over 50 independent runs (standard deviation is given after $\pm$). The best result is indicated in bold}
\label{dens}
\end{table}

\begin{table}
\centering
\begin{tabular}{ c c c c c c}
\hline
 Problem (b)  & bGP & bGP-I & bGP-P & GP & DM \\ 
 \hline
 $R^2$ & \bf{96.8 $\pm$ 1.34} & 95.9 $\pm$ 1.90 & 96.3 $\pm$ 1.58 & 95.6 $\pm$ 1.98  & 96.4 $\pm$ 1.54 \\
 $RMSE$ & \bf{0.58 $\pm$ 0.12} & 0.64 $\pm$ 0.14 & 0.61 $\pm$ 0.13 & 0.65 $\pm$ 0.14 & 0.60 $\pm$ 0.13 \\
 $CP (\%)$ & \bf{78.2 $\pm$ 16.8} & \bf{78.2 $\pm$ 16.8} & 74.4 $\pm$ 8.69 & 74.4 $\pm$ 8.69 & N/A \\
\hline
\end{tabular}
\caption{$R^2$, $RMSE$ (both value multiplied by 100) and $CP(\%)$ of examined methods for Problem (b). The reported values are the average over 50 independent runs (standard deviation is given after $\pm$). The best result is indicated in bold}
\label{squeeze}
\end{table}

\begin{table}
\centering
\begin{tabular}{ c c c c c c}
\hline
 Problem (c)  & bGP & bGP-I & bGP-P & GP & DM \\ 
 \hline
 $R^2$ & \bf{88.0 $\pm$ 3.51} & 69.1 $\pm$ 10.43 & 82.5 $\pm$ 7.74 & 65.4 $\pm$ 11.8  & 85.4 $\pm$ 6.20 \\
 $RMSE$ & \bf{1.68 $\pm$ 0.22} & 2.72 $\pm$ 0.43 & 1.99 $\pm$ 0.44 & 2.85 $\pm$ 0.49 & 1.83 $\pm$ 0.38 \\
 $CP (\%)$ & \bf{91.7 $\pm$ 11.5} & \bf{91.7 $\pm$ 11.5} & 88.3 $\pm$ 13.1 & 88.3 $\pm$ 13.1 & N/A \\
\hline
\end{tabular}
\caption{$R^2$, $RMSE$ (both value multiplied by 100) and $CP(\%)$ of examined methods for Problem (c). The reported values are the average over 50 independent runs (standard deviation is given after $\pm$). The best result is indicated in bold}
\label{daviga}
\end{table}

Table \ref{dens}-\ref{daviga} show that the full bGP approach yields highest $R^2$ and smallest $RMSE$ in all problems. Examining the performance between bGP and bGP-I (only inference) approach, as well as between the bGP-I and GP, it is immediate that the prediction accuracy benefits from the projection, as the approaches with projection always outperform their variants without projection regarding to $R^2$ and $RMSE$. The tables also show that the inference strategy incorporating bounds information is alway helpful, since both approaches with the bounded inference strategy (bGP and bGP-I) yields higher $R^2$ and lower $RMSE$ than their counterparts (bGP-P and GP). The performance gap, however, is smaller than the ones with and without projection, indicating the projection is generally more influential. No clear conclusions can be drawn from the comparison between bGP-P and DM, thus neither one can be favored over the other between the projection formula in Eq.\eqref{predmean} and the correlation-free conditional expectation formula in \cite{da2012gaussian}, if the posterior distribution functions are not of user's interest. Finally, with respect to the $95\%$ coverage probability $CP$, the GP parameters incorporating bound information lead to a better coverage (closer to nominal one $95\%$) in all problem except the problem (a). Overall, the results illustrate the added value brought by both the proposed projection scheme and the parameter inference scheme.

\subsubsection{2-D example}
\begin{figure}
\centering
\includegraphics[width=40pc]{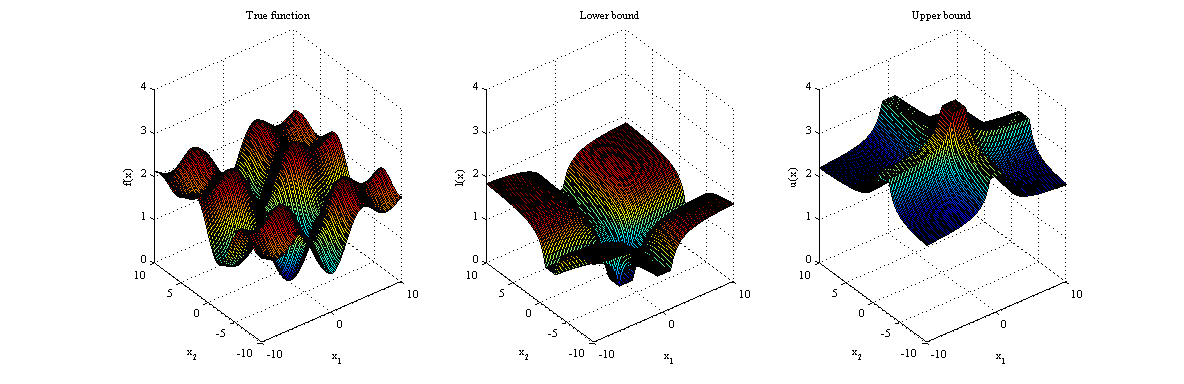}	
\caption{Response}
\label{rare_fig}
\end{figure}

We study a 2-dimensional function defined on box-bounded support $\mathcal{X}=[-10,10]\times[-10,10]$ as in \cite{auffray2014bounding}, which has several local minima and is not monotonic:
\begin{align}
f(\mathbf{x}) = -\frac{\sin (x_1)}{x_1}-\frac{\sin(x_2+2)}{x_2+2}+2
\end{align}  
and the following bound functions:
\begin{align*}
l(\mathbf{x}) =& \max [\min(-\frac{1}{x_1},\frac{1}{x_1}),-1] + \max [\min(-\frac{1}{x_2+2},\frac{1}{x_2+2}),-1] + 2 \\
u(\mathbf{x}) =& \min [\max(-\frac{1}{x_1},\frac{1}{x_1}),1] + \min [\max(-\frac{1}{x_2+2},\frac{1}{x_2+2}),1] + 2
\end{align*}
The landscape of all functions are shown in Figure \ref{rare_fig}.

In this problem, we examine the performance under different sizes of training data ($N$), ranging from 30 to 50. For each examined $N$, 50 independent simulations are replicated with $N$ LHS training samples and $N_t=1000$ uniformly distributed test samples in the input domain. The results are reported in Table \ref{auffray}.

We can observe that the full bGP approach performs the best for all examined training data size $N$. Both bGP-P and DM appear to work fairly well too. Similar to 1-D problems, the projection scheme greatly improves the prediction accuracy. More interestingly, accounting for the bounds in the parameter inference stage seem to be only beneficial when coupled with the projection scheme, as bGP is superior to bGP-P, but bGP-I gets outperformed by GP. This is explainable, as the criterion for parameter inference aims at minimizing the after-projection PRESS, which does not purposefully focus on the prediction performance without the projection scheme. For this problem, there is an slight improvement between DM and bGP-P. The GP parameters incorporating bound information lead to a better coverage (closer to nominal one $95\%$).

\begin{table}
\centering
\begin{tabular}{c c c c c c c}
\hline
 & $N$  & bGP & bGP-I & bGP-P & GP & DM \\
 \hline
 \multirow{3}{*}{$R^2$}&30 & \bf{80.5 $\pm$ 10.2} & 64.3 $\pm$ 17.5 & 77.9 $\pm$ 12.1 & 69.1 $\pm$ 15.8  & 80.2 $\pm$ 11.1 \\
 &40 & \bf{92.8 $\pm$ 4.17} & 86.1 $\pm$ 6.90 & 92.0 $\pm$ 3.91 & 89.3 $\pm$ 4.90 & 92.2 $\pm$ 3.61 \\
 &50 & \bf{97.2 $\pm$ 1.51} & 94.8 $\pm$ 3.63 & 96.6 $\pm$ 1.78 & 94.8 $\pm$ 3.31 & 96.8 $\pm$ 1.68 \\ 
 \hline
 \multirow{3}{*}{RMSE}&30 & \bf{21.8 $\pm$ 5.20} & 29.5 $\pm$ 6.83 & 23.1 $\pm$ 6.02 & 27.9 $\pm$ 6.65 & 22.1 $\pm$ 5.99 \\
 &40 & \bf{13.2 $\pm$ 3.70} & 18.5 $\pm$ 4.50 & 14.0 $\pm$ 3.50 & 16.3 $\pm$ 3.76 & 13.7 $\pm$ 3.82 \\
 &50 & \bf{8.29 $\pm$ 2.22} & 11.2 $\pm$ 3.43 & 9.13 $\pm$ 2.32 & 11.1 $\pm$ 3.07 & 8.70 $\pm$ 2.30 \\
 \hline
 \multirow{3}{*}{$CP$ ($\%$)}&30 & \bf{92.2 $\pm$ 11.4} & \bf{92.2 $\pm$ 11.4} & 88.2 $\pm$ 14.7 & 88.2 $\pm$ 14.7 & N/A \\
 &40 & \bf{94.8 $\pm$ 11.7} & \bf{94.8 $\pm$ 11.7} & 94.0 $\pm$8.72 & 94.0 $\pm$ 8.72 & N/A \\
 &50 & \bf{96.0 $\pm$ 5.30} & \bf{96.0 $\pm$ 5.30} & 93.8 $\pm$ 8.29 & 93.8 $\pm$ 8.29 & N/A  \\
 \hline
\end{tabular}
\caption{$R^2$ (multiplied by 100), RMSE and $CP$ ($\%$) of examined methods with different training data size $N$ in the 2-D example. The reported values are the average over 50 independent runs (standard deviation is given after $\pm$). The best result is indicated in bold}
\label{auffray}
\end{table}

\subsubsection{3-D example}
The Ishigami function is a classical 3-dimensional benchmark problem for sensitivity analysis \cite{saltelli2000sensitivity}. Each of its three input parameter belongs to the interval $[-\pi,\pi]$, and the Ishigami function is defined as:
\begin{align}
f(\mathbf{x}) = \sin(x_1)+7\sin^2(x_2)+0.1x_3^4\sin(x_1)
\end{align}
We adopt the Ishigami function as the output function to approximate, and assume the following bound functions:
\begin{align*}
l(\mathbf{x}) =& \max \left[\min(x_1,0),-1\right]\cdot (1+0.1x_3^4)\\
u(\mathbf{x}) =& \min \left[\max(x_1,0),1\right]\cdot (1+0.1x_3^4)+7\cdot \min(x_2^2,1)
\end{align*}
Both bounds are combinations of polynomial and min/max functions, and do not include the sine functions which creates the global nonlinear relationship in the output function.

We repeat 50 independent simulation trials for different values of $N$ from 20 to 100. The test samples are $N_t=1000$ uniformly distributed samples in the input domain. The results are reported in Table \ref{3d}.

The results in this problem display a nearly identical behavior as the previous 2-D example: bGP yields the best result when $N$ is small (20 to 40). As $N$ increases, the performance among bGP, bGP-P and DM become very similar. The projection is shown to be always beneficial, while the bound-incorporated parameter inference scheme improves the prediction quality when being coupled with the projection scheme. The GP parameters incorporating bound information lead to a better coverage (closer to nominal one $95\%$) in this example.

\begin{table}
\centering
\begin{tabular}{c c c c c c c}
\hline
 & $N$  & bGP & bGP-I & bGP-P & GP & DM \\ 
  \hline
\multirow{5}{*}{$R^2$} & 20 & \bf{50.5 $\pm$ 10.8} & 15.9 $\pm$ 26.1 & 43.2 $\pm$ 11.3 & 23.8 $\pm$ 20.8 & 47.4 $\pm$ 9.61 \\
& 40 & \bf{68.1 $\pm$ 8.57} & 40.1 $\pm$ 34.1 & 62.3 $\pm$ 11.4 & 48.1 $\pm$ 17.1 & 65.4 $\pm$ 10.1 \\
& 60 & 76.5 $\pm$ 6.72 & 51.4 $\pm$ 34.5 & 76.5 $\pm$ 7.42 & 68.5 $\pm$ 11.3 & \bf{78.1 $\pm$ 6.29} \\
& 80 & \bf{86.5 $\pm$ 4.37} & 77.5 $\pm$ 10.7 & 84.1 $\pm$ 5.46 & 79.5 $\pm$ 6.21 & 84.2 $\pm$ 5.08 \\
& 100 & 90.8 $\pm$ 4.29 & 85.4 $\pm$ 10.1 & 90.9 $\pm$ 2.96 & 86.9 $\pm$ 7.28 & \bf{91.3 $\pm$ 2.38} \\
 \hline
\multirow{5}{*}{RMSE} & 20 & \bf{2.61 $\pm$ 0.29} & 3.37 $\pm$ 0.52 & 2.79 $\pm$ 0.28 & 3.21 $\pm$ 0.43 & 2.65 $\pm$ 0.26 \\
& 40 & \bf{2.07 $\pm$ 0.28} & 2.80 $\pm$ 0.71 & 2.25 $\pm$ 0.35 & 2.64 $\pm$ 0.44 & 2.15 $\pm$ 0.32 \\
& 60 & 1.79 $\pm$ 0.26 & 2.49 $\pm$ 0.76 & 1.79 $\pm$ 0.27 & 2.07 $\pm$ 0.33 & \bf{1.73 $\pm$ 0.24} \\
& 80 & \bf{1.36 $\pm$ 0.22} & 1.74 $\pm$ 0.34 & 1.47 $\pm$ 0.25 & 1.68 $\pm$ 0.25 & 1.44 $\pm$ 0.24 \\
& 100 & 1.11 $\pm$ 0.24 & 1.37 $\pm$ 0.42 & 1.11 $\pm$ 0.18 & 1.33 $\pm$ 0.33 & \bf{1.08 $\pm$ 0.15} \\
 \hline
\multirow{5}{*}{$CP$ ($\%$)} & 20 & \bf{89.7 $\pm$ 16.3} & \bf{89.7 $\pm$ 16.3} & 78.6 $\pm$ 14.8 & 78.6 $\pm$ 14.8 & N/A \\
& 40 & \bf{89.9 $\pm$ 18.2} & \bf{89.9 $\pm$ 18.2} & 82.1 $\pm$ 7.71 & 82.1 $\pm$ 7.71 & N/A \\
& 60 & \bf{90.0 $\pm$ 15.5} & \bf{90.0 $\pm$ 15.5} & 85.4 $\pm$ 8.06 & 85.4 $\pm$ 8.06 & N/A  \\
& 80 & \bf{91.2 $\pm$ 13.6} & \bf{91.2 $\pm$ 13.6} & 86.8 $\pm$ 7.77 & 86.8 $\pm$ 7.77 & N/A \\
& 100 & \bf{94.0 $\pm$ 7.38} & \bf{94.0 $\pm$ 7.38} & 91.4 $\pm$ 19.8 & 91.4 $\pm$ 19.8 & N/A \\
 \hline
\end{tabular}
\caption{$R^2$ (multiplied by 100), RMSE and $CP$ ($\%$) of examined methods with different training data size $N$ in the 3-D example. The reported values are the average over 50 independent runs (standard deviation is given after $\pm$). The best result is indicated in bold}
\label{3d}
\end{table}

\subsection{A real-data example with multi-fidelity bounds}

Let us consider the maximum likelihood inference task in Astrophysics \cite{kandasamy2016gaussian, davis2007scrutinizing, zaytsev2017minimax}. The input consists of three cosmological parameters: the Hubble Constant $H_0 \in (60, 80)$, Dark Matter
Fraction $\Omega_M \in (0, 1)$ and Dark Energy Fraction $\Omega_{\Lambda} \in (0,1)$. The likelihood is computed using the Robertson-Walker metric, which models the distance to a supernova given the input parameters and the observed red-shift data (taken from \cite{davis2007scrutinizing}). Such likelihood evaluation cannot be analytically conducted, and requires one dimensional numerical integration for each data point. 

For this problem, a multi-fidelity data set is available \cite{zaytsev2017minimax}, containing 5000 input points and their corresponding high-fidelity likelihood evaluation $l_h$ (for which the one-dimensional integrations are performed using the trapezoidal rule on a grid of size 1000) and low-fidelity ones $l_l$ (grids of size 3). We choose the output function $f(\mathbf{x})$ to be high-fidelity likelihoods, and utilize the cheap low-fidelity ones to establish its bound, representing the real-world scenario where the output evaluations are expensive but their associated bounds can be obtained in a much lower cost. Specifically, the upper bound $u(\mathbf{x})$ is expressed as follows:
\begin{align*}
f(\mathbf{x})=l_h(\mathbf{x})\leq u(\mathbf{x}) = l_l(\mathbf{x})+8
\end{align*}
Such upper bounds have been tested to hold for all 5000 inputs in the data set.

We will examine different sizes of $N$ ranging from 10 to 30. For each $N$, 50 independent simulation trials are repeated. In each simulation, the whole data set is randomly divided into a training set of size $N$ and a cross-validation set of size ($N_t=5000-N$). The accuracy measures $R^2$, $RMSE$ and $CP$ are evaluated based on the cross-validation set and listed in Table \ref{real-world}. 

Overall the trends observed from the synthetic examples are again confirmed for this real-data problem. bGP outperform other alternatives in most cases. The benefit of the projection is significant and particularly noticeable when the training set is small. In contrast to the 2-D and 3-D problems, the GP parameter inference strategy incorporating bounds is shown to increase the prediction accuracy, no matter whether it is combined with or without the projection scheme. The performance gap between bGP-P and DM is once again not significant in this example. For this problem, the coverage probabilities are significantly narrower than the nominal one ($95\%$), but GP parameters incorporating bound information lead to a broader coverage for most cases in this example.

\begin{table}
\centering
\begin{tabular}{c c c c c c c}
\hline
  & $N$  & bGP & bGP-I & bGP-P & GP & DM \\ 
 \hline
\multirow{5}{*}{$R^2$}&  10 & \bf{97.7 $\pm$ 2.80} & 92.2 $\pm$ 15.5 & 97.2 $\pm$ 4.16 & 89.7 $\pm$ 16.5 & 96.9 $\pm$ 4.30 \\
& 15 & \bf{99.5 $\pm$ 0.64} & 98.9 $\pm$ 1.33 & 99.2 $\pm$ 1.69 & 98.3 $\pm$ 2.70 & 99.2 $\pm$ 1.69 \\
& 20 & \bf{99.7 $\pm$ 0.32} & 99.3 $\pm$ 0.88 & 99.6 $\pm$ 0.48 & 99.0 $\pm$ 1.78 & 99.5 $\pm$ 0.43 \\
& 25 & \bf{99.9 $\pm$ 0.25} & 99.8 $\pm$ 0.25 & 99.8 $\pm$ 0.06 & 99.7 $\pm$ 0.10 & 99.7 $\pm$ 0.06 \\
& 30 & 99.9 $\pm$ 0.07 & 99.9 $\pm$ 0.16 & 99.9 $\pm$ 0.06 & 99.9 $\pm$ 0.06 & \bf{99.9 $\pm$ 0.05} \\
 \hline
\multirow{5}{*}{RMSE}& 10 & \bf{36.0 $\pm$ 22.8} & 65.7 $\pm$ 52.1 & 39.0 $\pm$ 28.2 & 71.5 $\pm$ 55.7 & 41.1 $\pm$ 30.2 \\
& 15 & \bf{15.9 $\pm$ 9.12} & 25.5 $\pm$ 10.9 & 19.7 $\pm$ 15.8 & 29.8 $\pm$ 19.0 & 19.7 $\pm$ 16.1 \\
& 20 & \bf{12.6 $\pm$ 6.80} & 18.8 $\pm$ 9.98 & 14.0 $\pm$ 8.60 & 21.4 $\pm$ 14.0 & 14.1 $\pm$ 8.49 \\
& 25 & \bf{8.47 $\pm$ 5.78} & 11.1 $\pm$ 5.14 & 8.99 $\pm$ 8.32 & 11.5 $\pm$ 9.32 & 8.90 $\pm$ 8.29 \\
& 30 & 5.69 $\pm$ 2.90 & 7.27 $\pm$ 2.83 & 5.36 $\pm$ 2.64 & 6.99 $\pm$ 2.58 & \bf{5.36 $\pm$ 2.74} \\
\hline
\multirow{5}{*}{$CP$ ($\%$)}& 10 & 48.7 $\pm$ 30.9 & 48.7 $\pm$ 30.9 & \bf{56.2 $\pm$ 23.3} & \bf{56.2 $\pm$ 23.3} & N/A \\
& 15 & \bf{59.6 $\pm$ 25.0} & \bf{59.5 $\pm$ 25.0} & 55.3 $\pm$ 24.3 & 55.3 $\pm$ 24.3 & N/A \\
& 20 & \bf{65.9 $\pm$ 28.1} & \bf{65.9 $\pm$ 28.1} & 58.9 $\pm$ 22.3 & 58.9 $\pm$ 22.3 & N/A  \\
& 25 & \bf{70.8 $\pm$ 25.8} & \bf{70.8 $\pm$ 25.8} & 60.7 $\pm$ 23.5 & 60.7 $\pm$ 23.5 & N/A \\
& 30 & \bf{72.8 $\pm$ 26.3} & \bf{72.8 $\pm$ 26.3} & 62.5 $\pm$ 21.5 & 62.5 $\pm$ 21.5 & N/A \\
 \hline
\end{tabular}
\caption{$R^2$ (multiplied by 100), RMSE and $CP$ ($\%$) of examined methods with different training data size $N$ in the real-data example. The reported values are the average over 50 independent runs (standard deviation is given after $\pm$). The best result is indicated in bold}
\label{real-world}
\end{table}

\section{Application to probability density approximation}

A fundamental problem in modern statistics society is to approximate difficult-to-compute probability densities \cite{blei2017variational}. A good density approximation can be utmost important to the subsequent Bayesian inference and/or the construction of a good proposal distribution for advanced Monte Carlo simulations (\eg importance sampling  or independent Metropolis-Hastings \cite{giordani2010adaptive}). Consequently, the field of density approximation has received a vast amount of interest in recent years. Relevant approaches includes \emph{Laplace approximation}, which approximates the density by a normal distribution around the density mode \cite{bornkamp2011approximating}; \emph{Variational Inference}, which finds the closest member to the target density from a pre-defined family of densities; and various sampling-based approaches that approximate the density by empirical estimates based on collected samples from that density.

In \cite{joseph2012bayesian}, an appealing method named \emph{DoIt} was introduced, with the core idea of approximates or interpolates the continuous target distributions using GP. This approach has been shown to be particularly useful for approximating computationally expensive densities. However as acknowledged by the author, an apparent drawback of the original DoIt is the lack of guarantee about the GP interpolation to be positive everywhere. Recently a modification of DoIt is proposed to guarantee the non-negativity by interpolating the squared root of the density \cite{joseph2013note}.

\begin{figure}
\centering
\includegraphics[width=30pc]{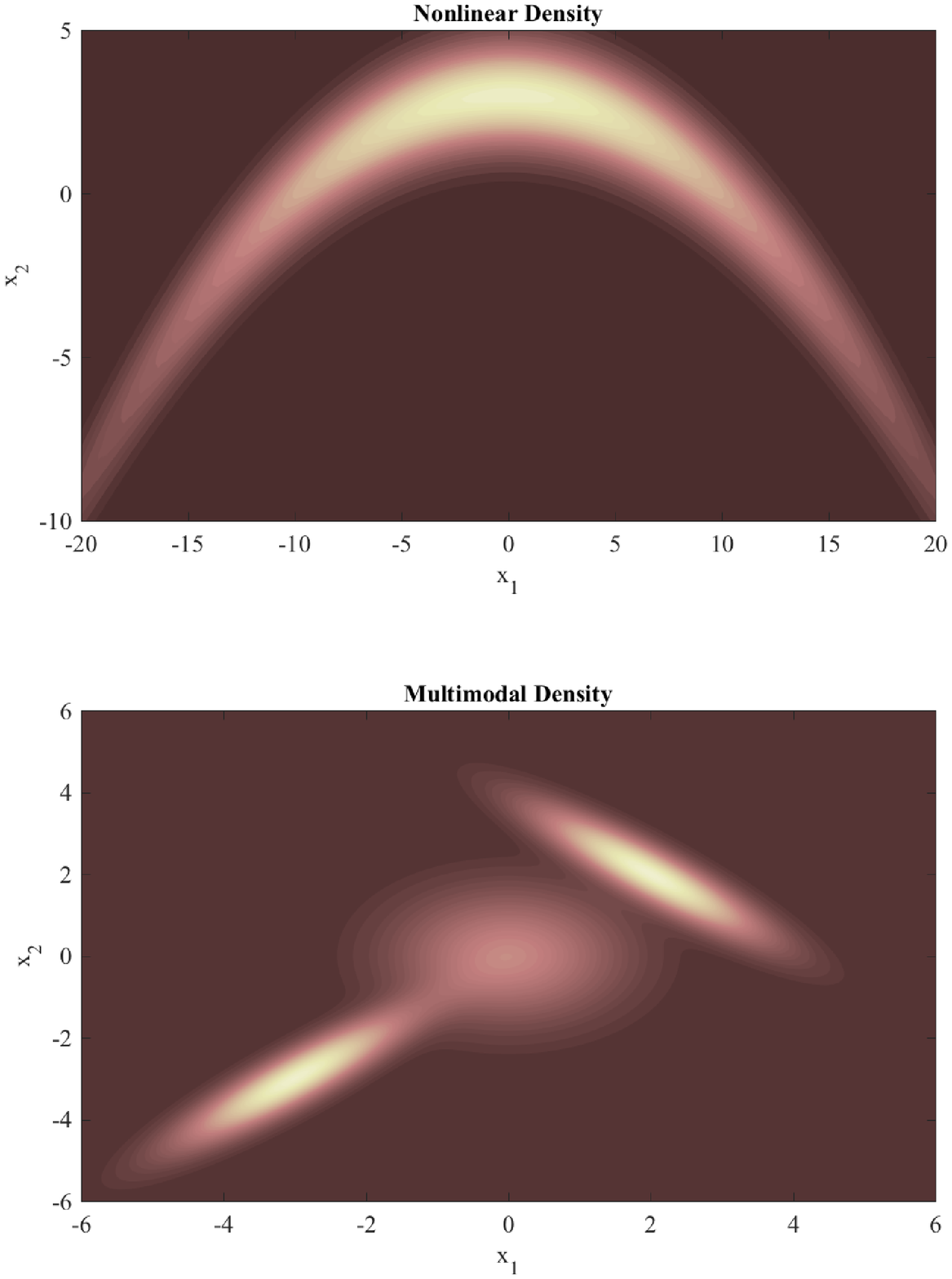}	
\caption{Contour plots of target densities}
\label{densities}
\end{figure}

Along the spirit of DoIt, we consider the density approximation problem as a non-negative interpolation problem, where the output for GP is taken as the density to be approximated. Only the constant lower bound $l(\mathbf{x})=0$ is adopted to enforce non-negative approximation. We consider to approximate two challenging densities, with strong nonlinearity or multimodality features. The first density is a twisted Gaussian distribution with banana-shaped contours (as can be seen on the upper part of Figure \ref{densities}) discussed in \cite{haario1999adaptive, joseph2012bayesian}:
\begin{align*}
f(\mathbf{x}) = \phi([x_1,x_2+0.03x_1^2-3]^T;[0,0]^T,\left[\begin{matrix}100 & 0\\ 0 & 100\end{matrix}\right])
\end{align*}
This density exhibits high nonlinearity, and will be referred as the \emph{nonlinear} density.

The second density is a Gaussian mixture density as used in \cite{gilks1998adaptive, bornkamp2011approximating}, whose PDF is expressed as follows:
\begin{align*}
f(\mathbf{x}) = 0.34\cdot\phi(\mathbf{x};[0,0]^T,\left[\begin{matrix}1 & 0\\ 0 & 1\end{matrix}\right])+0.33\cdot\phi(\mathbf{x};[-3,-3]^T,\left[\begin{matrix}1 & 0.9\\ 0.9 & 1\end{matrix}\right])+0.33\cdot\phi(\mathbf{x};[2,2]^T,\left[\begin{matrix}1 & -0.9\\ -0.9 & 1\end{matrix}\right])
\end{align*}
This density is naturally multimodal, and displays a complicated local structure; see the lower part of Figure \ref{densities}. It will be further referred as the \emph{multimodal} density. 

Now suppose that the target density values at $N$ input locations are obtained. One can then construct a GP on such observations, and approximate the density by the GP predictor $\hat{f}$ at any arbitrary input. The predictor $\hat{f}$ represents unnormalized approximations, in the sense that it does not integrate up to one. To transform $\hat{f}$ to a valid density integrating up to one, we can normalize such density approximation as follows: 
\begin{align*}
\bar{f}(\mathbf{x}) = \frac{\hat{f}(\mathbf{x})}{\hat{F}} = \frac{\hat{f}(\mathbf{x})}{\int_{\mathcal{X}} \hat{f}(\mathbf{x})d\mathbf{x}}
\end{align*}
where $\hat{F}$ is the normalization constant that has to be numerically computed, and $\bar{f}$ is the normalized density approximation.

In this application, due to the non-negative requirement on GP predictors, we only implement the three applicable approaches, namely the Full, Projection, and DM. It is of immediate interest to see how the density approximation from different approaches compare with the true density. For this purpose, we employ the squared Hellinger distance $H^2(f,\bar{f})$ to measure the discrepancy between the true density $f$ and the approximated density $\bar{f}$: 
\begin{align}
H^2(f,\bar{f}) = \int_{\mathcal{X}} \frac{1}{2}\left(\sqrt{f(\mathbf{x})}-\sqrt{\bar{f}(\mathbf{x})}\right)^2d\mathbf{x}
\label{hel}
\end{align}
Such distance is bounded in the range of [0, 1], with larger values corresponding to higher discrepancies and worse approximations.

For the numerical implementation, the following setup is adopted. In order to reduce the potential influence of extrapolation by restricting the support of target densities into a box-bounded region (specifically, $\mathcal{X}$ is set as $[-20,20]\times[-10,5]$ for nonlinear, and $[-6,6]\times[-6,6]$ for multimodal). The training inputs are chosen as LHS samples in such box-bounded region. All numerical integrations required for obtaining the normalization constants or the squared Hellinger distances have been calculated by the means of Monte Carlo simulations (MCS), with $10^6$ simulation samples from a uniform distribution over the box-bounded region. The estimation error from MCS has been examined to be negligible under such a sample size. 50 independent simulation trials are repeated for different values of $N$ from 50 to 500. The squared Hellinger distances under all approaches are reported in Table \ref{h2}.  

\begin{table}
\centering
\begin{tabular}{ c c c c| c c c }
\hline
    & \multicolumn{3}{c}{Nonlinear} & \multicolumn{3}{c}{Multimodal}\\
\hline
  $N$ & Full & Projection & DM & Full & Projection & DM\\ 
 \hline
 50 & \bf{0.173 $\pm$ 0.041} & 0.218 $\pm$ 0.047 & 0.263 $\pm$ 0.044 & \bf{0.272 $\pm$ 0.051} & 0.299 $\pm$ 0.046 & 0.339 $\pm$ 0.043 \\
 100 & \bf{0.088 $\pm$ 0.012} & 0.124 $\pm$ 0.017 & 0.171 $\pm$ 0.017 & \bf{0.185 $\pm$ 0.029} & 0.247 $\pm$ 0.028 & 0.311 $\pm$ 0.028 \\
 200 & \bf{0.041 $\pm$ 0.005} & 0.061 $\pm$ 0.006 & 0.093 $\pm$ 0.009 & \bf{0.106 $\pm$ 0.020} & 0.153 $\pm$ 0.017 & 0.215 $\pm$ 0.021 \\
 500 & \bf{0.008 $\pm$ 0.001} & 0.012 $\pm$ 0.002 & 0.021 $\pm$ 0.003 & \bf{0.031 $\pm$ 0.005} & 0.055 $\pm$ 0.005 & 0.090 $\pm$ 0.009 \\
 \hline
\end{tabular}
\caption{$H^2$ of the 3 examined methods with different training data size $N$. The results are averaged based on 50 simulation replicates. Standard deviation is given after $\pm$}
\label{h2}
\end{table}

From Table \ref{h2}, one can easily conclude that the Full approach substantially improves the density approximation quality in terms of the squared Hellinger distance by a clear margin. Its consistently superior performance should be attributed to incorporating bound information in parameter inference. In addition, the Projection approach also gives better results than DM. As both approaches uses identical GP parameters, the performance discrepancy indicates the projection in Eq. \eqref{predmean} helps yielding better density approximation than the correlation-free formulation in \cite{da2012gaussian}.

\begin{figure}
\centering
\includegraphics[width=40pc]{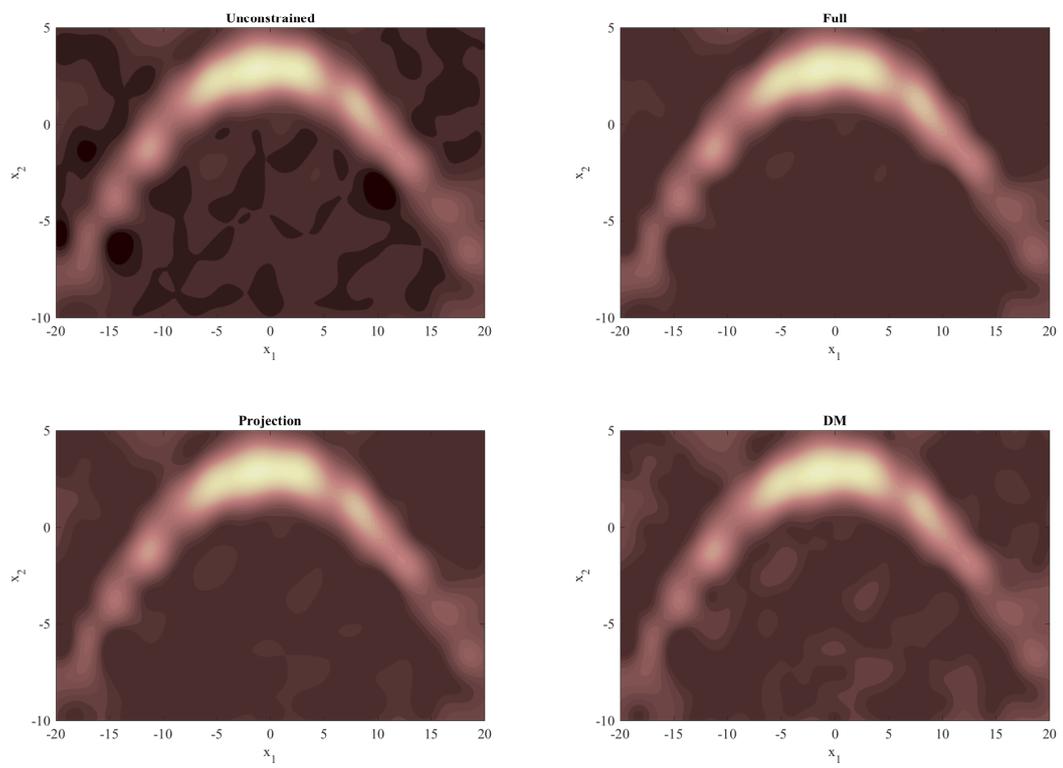}
\caption{Contour plots of unnormalized density approximations by different approaches}
\label{approx_densities}
\end{figure}

To graphically compare performances, in Figure \ref{approx_densities}, we present contour plots of the unnormalized approximation ($\hat{f}$) for the nonlinear case by all compared approaches with the same set of $N=200$ experiments. The prediction from the Unconstrained approach is also plotted. Similar to the quantitative results above, the Full approach again gives the best approximation to the true target density visually as shown in Figure \ref{densities}, especially in low-density regions (\ie darker regions in the contour plot). On the contrary, the Unconstrained approach suffers heavily from capturing the shape of low-density regions. This comparison again emphasizes the importance of accounting for the bound information. 

\section{Discussion}\label{sec:con}
 In this paper, we developed a methodology for the incorporation of bound information into the Gaussian processes. Based on the projection methodology, GP sample paths can be projected to honor bounds across the entire input domain. The closed-formed solution for the projection is proposed. Notably, the projected posterior distribution at any input will be a mixed discrete-continuous distributions, consisting of a degenerated two-point distribution at bounds and a truncated Gaussian distribution within the bounds. For such projected GP posterior distribution, the closed-formed formulas of the mean, variance, and distribution functions are proposed. A strategy to infer the GP parameters considering the bound information is proposed. It projects the leave-one-out posterior density to obtain the LOO-CV estimator, and obtains the GP parameters by minimizing the sum of squared errors between the projected estimator and true response functions. In addition, a constraint is proposed to further enhance the numerical stability of such strategy. Both the projection as well as the parameter inference scheme can be easily implemented with a comparable computational complexity with respect to the unconstrained one. Finally, through synthetic examples from 1D to 3D, a real-data multi-fidelity problem, as well as applications to density approximation, we show that it is beneficial to apply both the projection technique and the bound-incorporated GP parameter inference strategy.
 
 Despite that only the zero-mean Gaussian processes with the squared exponential covariance function is adopted in all examples, the proposed methods are easily applied to other mean and covariance functions as long as the GP sample paths are continuous, \ie Universal Kriging with Generalized Exponential covariances. In addition, for some applications (\eg the problem (c)), the continuous bound requirement may not be met. We also demonstrate that one may simply treat the proposed method as a heuristic correction formula for such problems, and the results are also encouraging.

\section*{Acknowledgement}

The contribution of LL is funded by NSFgrants IIS 1663870 and DMS CAREER 1654579.

\section*{References}

\bibliography{JZ}

\newpage
\appendix
%
%
%
%

\end{document}